\documentclass[
    aps,
    prd,
    twocolumn,
    nofootinbib,
    superscriptaddress,
    amsmath,
    amssymb
]{revtex4-2}

\usepackage{graphicx}
\usepackage{bm}
\usepackage{ytableau}
\usepackage{slashed}
\usepackage{physics}

\usepackage{appendix}

\begin{document}

\title{Higher Spin Lagrangians for Black Holes in Arbitrary Dimension}

\author{Francesco Campanella}
\email{francesco.campanella@uniroma1.it}
\affiliation{Dipartimento di Fisica, Sapienza
Università di Roma, Piazzale Aldo Moro 2,
00185 Roma, Italy}


\begin{abstract}
In this paper, we investigate the classical gravitational multipolar structure of rotating objects studying higher-spin Lagrangians coupled to gravity in arbitrary dimensions $D=d+1$. We first consider fields described by totally symmetric tensors with $s$ indices. At finite spin, the minimal coupling depends on both the spin and spacetime dimension. We show that, in the high-spin limit $s\gg 1$, the minimal multipolar expansion becomes dimension independent and it corresponds to the one which reproduces the Kerr solution in four dimensions. Furthermore, we identify dimension independent relations among the non-minimal coupling coefficients required to reproduce the Myers-Perry multipolar structure in arbitrary dimensions. We then extend our analysis to Lagrangians involving fields with different Young representations, which are allowed in higher dimensions.  
\end{abstract}

\maketitle

\tableofcontents

\section{Introduction}
Black holes (BHs) are among the most fascinating and mysterious objects in the Universe. Predicted by General Relativity (GR), they have been the subject of extensive theoretical investigation over the past century, aimed at understanding their fundamental properties. Among them, uncharged rotating black holes are of particular phenomenological interest. In four dimensions, the only solution is the Kerr black hole \cite{Kerr:1963ud}. In higher dimensions, however, the horizon topology can be more intricate, allowing for a richer spectrum of solutions. These range from the Myers-Perry black holes  \cite{Myers:1986un}, which generalize the Kerr solution to arbitrary dimensions, to more intricate objects such as Black Ring \cite{Emparan:2001wn} and Black Saturn \cite{Elvang:2007rd}.
Another important distinction between four- and higher-dimensional solutions lies in the multipolar structure. Four dimensional asymptotically flat vacuum solutions are characterized by two sets of multipole moments: mass and current multipoles \cite{Geroch:1970cc,Geroch:1970cd,Hansen:1974zz,Thorne:1980ru}. In higher dimensions an additional set of multipole arises, called "stress multipoles", as they are associated to the spatial part of the metric \cite{Gambino:2024uge,Bianchi:2024shc}.
In the last years, scattering amplitudes techniques have been successfully employed to study rotating black holes \cite{Vaidya:2014kza,Vines:2017hyw,Arkani-Hamed:2017jhn,Chung:2018kqs,Guevara:2018wpp,Guevara:2019fsj,Arkani-Hamed:2019ymq,Maybee:2019jus,Bern:2020buy,Aoude:2021oqj,Alessio:2023kgf,Cangemi:2026cjy, Monteiro:2026rby,Bianchi:2025xol} and the gravitational dynamics of spinning compact objects 
\cite{Kosmopoulos:2021zoq,Liu:2021zxr,Chen:2021kxt,Menezes:2022tcs,Bern:2022kto,Bautista:2023szu,Bohnenblust:2024hkw,Chen:2024mmm,Akpinar:2024meg,FebresCordero:2022jts,Jakobsen:2023ndj,Jakobsen:2022fcj,Damgaard:2022jem,Luna:2023uwd, Gonzo:2024zxo,Akpinar:2025huz,Alaverdian:2025jtw,Akpinar:2025bkt,Akpinar:2025byi,Gatica:2025uhx}. 
In this context, Ref. \cite{Chung:2018kqs} showed, implementing the spinor helicity formalism, that the three-point amplitude describing the emission of a graviton from a minimally coupled massive particle in four dimensions reproduces, in the infinite spin limit, the stress-energy tensor of a Kerr BH in momentum space \cite{Vines:2017hyw}. In particular, in the classical limit, a minimally coupled spin $s$ particle reproduces this stress-energy tensor up to order $2s$ in the angular momentum.
The connection between scattering amplitudes and the multipolar structure of rotating objects naturally motivates the use of massive higher-spin field theories.
The study of higher-spin has a long history, motivated by the aim of constructing consistent relativistic descriptions of particles with arbitrary spin. A first systematic formulation was developed by Singh and Hagen in 1974 for massive fields \cite{PhysRevD.9.898}, while massless description was realized by Fronsdal in 1978 \cite{PhysRevD.18.3624}. In his seminal work, Fronsdal constructed a gauge invariant Lagrangian formulation for massless fields with arbitrary integer spin, described by completely symmetric tensor fields, generalizing Maxwell's theory and linearized gravity. A further generalization was developed by Labastida in the late 1980s \cite{Labastida:1987kw}, extending the Fronsdal formulation to tensor fields with mixed-symmetry properties, classified by Young Tableau with multiple rows. Related works on massless mixed-symmetry fields can be found in Refs. \cite{Campoleoni_2009,Campoleoni:2009je}.
The construction of gauge invariant Lagrangians for massive higher-spin fields relies on the Stückelberg formalism  \cite{Stueckelberg:1938hvi}, which consists in introducing a set of lower rank auxiliary fields to compensate for the breaking of gauge invariance induced by the mass term. This formalism has been employed in the description of both higher-spin and mixed-symmetry fields \cite{zinoviev2002massivemixedsymmetrytensor,Zinoviev:2001dt,Zinoviev:2003ix,Zinoviev:2009gh, Reshetnyak:2023oyj, Buchbinder:2007ix, Buchbinder:2011xw,Reshetnyak:2026lno,Reshetnyak:2012ec}.
Within the framework of Effective Field Theories (EFTs), higher-spin fields have been employed to describe rotating objects in four dimensions. In Ref. \cite{Cangemi:2022bew}, the authors showed how the three-point Kerr amplitude is reproduced by minimally coupled higher-spin Lagrangians. These results have been used to study the Compton amplitude of root-Kerr and Kerr \cite{Cangemi:2023ysz,Cangemi:2023bpe, Bohnenblust:2024hkw}. Further developments in this direction can be found in Refs. \cite{Zinoviev:2025tpx,Skvortsov:2023jbn}. 
The present work is set in this context. Recently, a systematic procedure for extracting classical spin information from the amplitudes in arbitrary dimensions was developed in Ref. \cite{Campanella:2026wqt}. Furthermore, the authors demonstrated that, in higher dimensions, the minimal coupling depends both on the spacetime dimensionality and the field representation, and it does not reproduce anymore a well known gravitational multipolar structure.
From this, we asked whether a universal multipolar structure can be identified for minimally coupled massive fields in arbitrary dimensions. This paper addresses this question by considering large-spin symmetric and mixed-symmetry massive fields, whose representations are allowed in higher dimensional spacetimes. 

The paper is organized as follows.
In Section \ref{sez 1} we show the systematic procedure to extract the classical spin information from the amplitude.
In Section \ref{sez 2} we analyze the multipolar structure generated by totally symmetric fields Lagrangians both minimally and non-minimally coupled to gravity. 
In Section \ref{sez 3} we extend the analysis to objects with mixed symmetries, in particular we focus on fields with two families of $s$ and $n$ symmetric indices, which corresponds to two-rows Young Tableau.
Finally, Section \ref{sez 4} contains final discussion.
In the Appendix-\ref{Notations}, the notations used in the paper can be found.

\section{Spin-Tensor structures from the amplitude}
\label{sez 1}
\begin{figure}[ht]
    \centering
    \includegraphics[width=0.39\linewidth]{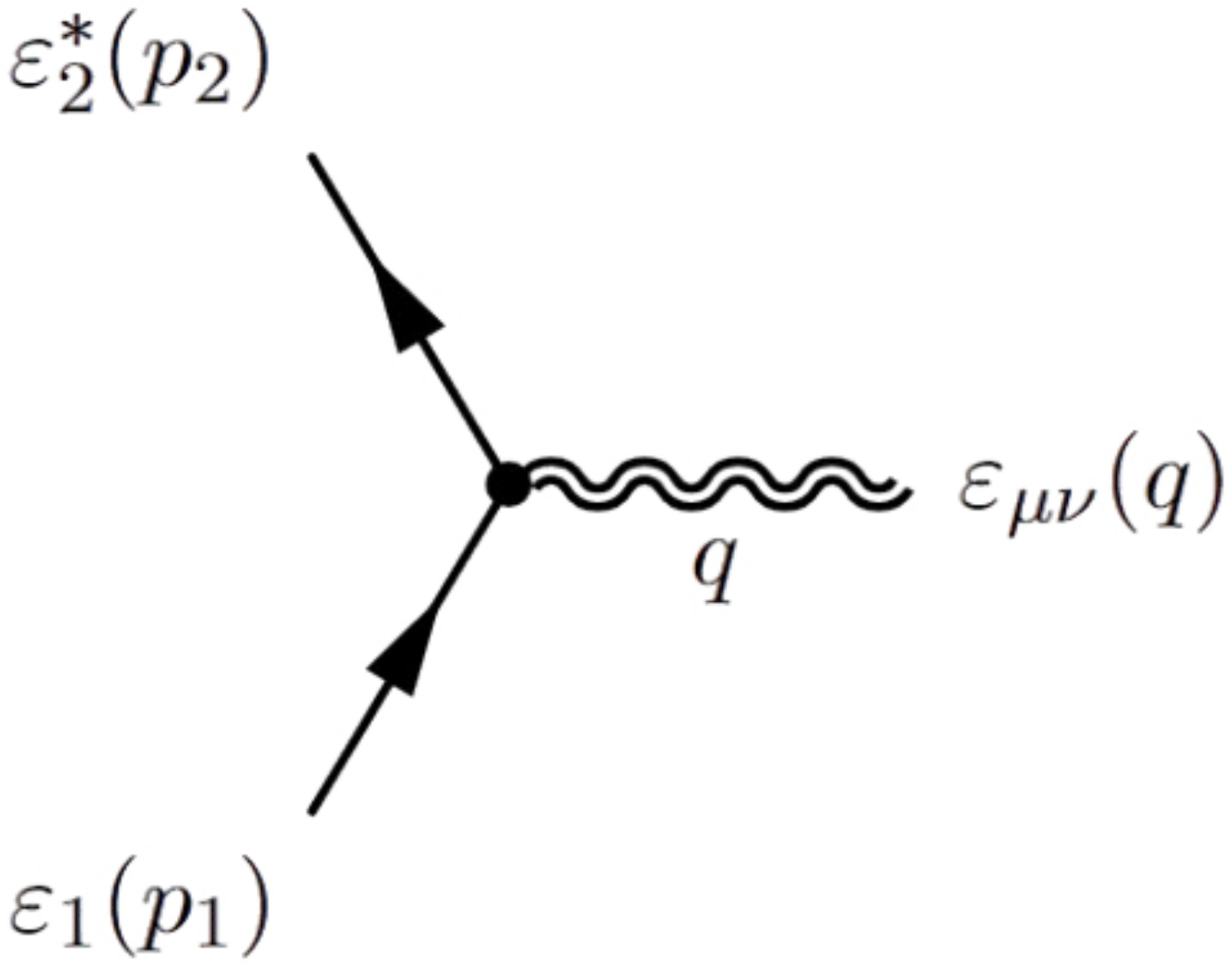}
    \caption{3-point amplitude involving spin massive fields and a graviton.}
    \label{tree}
\end{figure}

In this section, we review the methods presented in \cite{Campanella:2026wqt, Vaidya:2014kza} to extract the multipolar expansion from the 3-point function involving spinning fields emitting/absorbing a graviton.
One reads the stress-energy tensor from the amplitude in Figure \ref{tree}, where the
two massive particles have equal mass and spin and $q= p_1- p_2 $ is the transferred momentum. We can write the massive-particle momenta in terms of the center-of-mass momentum and the transferred momentum as
\begin{equation}
    p_1^\mu=P^\mu+\frac{q^\mu}{2}, \quad\quad p_2^\mu=P^\mu-\frac{q^\mu}{2} \ .
\label{momenti 2}
\end{equation}
The angular momentum is computed directly from the stress-energy tensor \cite{Bjerrum-Bohr:2002fji},  resulting in the identification
\begin{equation} \label{eq: SO(d)tensorfullycov}
    J^{\mu\nu}=\frac{\varepsilon^{B}_2(M^{\mu\nu})_B{}^{A}\varepsilon_1{}_{A}}{\varepsilon_2\cdot\varepsilon_1} \ ,
\end{equation}
where the indices $A$ and $B$ denote a generic, arbitrary spin representation for the polarization tensors, and $(M^{\mu\nu})_A{}^{ B}$ are the Lorentz generators in that representation, satisfying the algebra
\begin{small}
\begin{equation}
    [M^{\mu \nu}, M^{\rho \sigma}]=-i\Big(\eta^{\mu\rho}M^{\nu\sigma}-\eta^{\nu \rho}M^{\mu \sigma}+\eta^{\nu \sigma}M^{\mu \rho}-\eta^{\mu\sigma}M^{\nu\rho}\Big)\ .
    \label{Lorentz algebra}
\end{equation} 
\end{small}
In the expression above, the polarizations are the ones of the center-of-mass momentum, implying that in order to read the expansion in the spin tensor inside the amplitude, one has to boost them back by $\pm\frac{q}{2}$.
This is precisely what the Spin Supplementary Condition (SSC) takes care of  (see {\it e.g.} \cite{Guevara:2018wpp} and references therein). 

We can choose the frame to be the center-of-mass frame, so that the tensor in eq. \eqref{eq: SO(d)tensorfullycov} becomes the $SO(d)$ tensor with only spatial indices
\begin{equation} 
    J^{ij}=\frac{\varepsilon^{B}_2(M^{ij})_B{}^{A}\varepsilon_1{}_{A}}{\varepsilon_2\cdot\varepsilon_1} \ ,
    \label{SO(d) little group}
\end{equation}
where $(M^{ij})_A{}^{B}$ are the little group generators. This object automatically satisfies SSC.
The strategy to compute the angular-momentum dependence within a 3-point scattering amplitude is therefore to expand the polarizations around the static frame in powers of the transferred momentum \cite{Vaidya:2014kza}. In the following we are going to describe the procedure in detail.   

A generic theory describing a massive field coupled to gravity yields a 3-point amplitude which is a function of the massive momenta $p_1$ and $p_2$ and the transferred momentum. We first rewrite the amplitude in terms of $P$ and $q$, and we observe that in the center-of-mass frame the first is purely time-like while the second is purely space-like.
On-shellness of the graviton polarization $\epsilon_{\mu\nu}$ implies that terms proportional to $q^\mu$  and $q^\nu$, as well as terms proportional to $\eta^{\mu\nu}$, vanish. From the point of view of the classical theory, this corresponds to neglecting terms of the metric that can be gauged away within the gauge chosen \cite{Gambino:2024uge}, and therefore do not contribute to the multipolar expansion. We further neglect terms proportional to $q^2$, that would lead to local contributions to the metric. This implies that we can substitute $P^\mu$ with $m u^\mu$, where $u^\mu=\delta^\mu_0$ is the velocity of the source.
In the amplitude, the polarization tensors $\varepsilon_{1,2}(p_{1,2})$  are also expanded around the rest frame polarizations $\varepsilon_{1,2}$, and this gives in general terms of the form $\varepsilon_{1,A} \varepsilon_2^B$ in different irreducible representations of the little group $SO(d)$. Using the specific form of the generators $(M^{ij})_B{}^A$,  each irreducible representation $\mathcal{R}_{\mathcal{I}}$  can be uniquely written in terms of specific products of the generators $M$'s  projected on the same representation, as
\begin{widetext}
\begin{equation} 
   \underbrace{ \varepsilon_{2}^B \varepsilon_{1,A} }_{\mathcal{R}_{\mathcal{I}}}  = C_{\mathcal{R}_{\mathcal{I}}}(\underbrace{ M \cdot M \cdot \ldots  \cdot M }_{\mathcal{R}_{\mathcal{I}}})_A{}^B \bigg[\varepsilon_2^D(\underbrace{M \cdot M \cdot \ldots \cdot M}_{\mathcal{R}_{\mathcal{I}}})_D{}^C\varepsilon_{1,C}\bigg] \ .
    \label{Jirrep}
\end{equation}
\end{widetext}
Generalizing eq. \eqref{SO(d) little group}, one then  determines the various powers of the angular momentum in the amplitude through the identification
\begin{equation}
    \underbrace{J\cdot J \cdot \ldots\cdot  J}_{\mathcal{R}_{\mathcal{I}}}=\frac{1}{\varepsilon_2\cdot\varepsilon_1}\varepsilon_2^D(\underbrace{M\cdot M \cdot \ldots \cdot  M}_{\mathcal{R}_{\mathcal{I}}})_D{}^C\varepsilon_{1,C} \ ,
\end{equation}
where the number of generators involved in the representation $\mathcal{R}_{\mathcal{I}}$ gives the order of the multipole. Note that in general not all the representations that appear in the decomposition \eqref{Jirrep} correspond to actual multipoles. By consistency, these representations must disappear in the amplitude in the classical limit by symmetrisation when contracted with the momenta up to local or gauge terms.  

Once this analysis is carried out, so that the amplitude is written in terms of the spin tensor $J_{ij}$, one follows~\cite{Kosower:2018adc} to schematically obtain the classical limit out of scattering amplitude calculations. This is obtained making the replacement
\begin{equation}
    q\rightarrow \hbar\,  q \qquad \text{and} \qquad { J}\rightarrow\frac{1}{\hbar}{ J} \ , \label{eq:KMOC}
\end{equation}
keeping only the terms $O(\hbar^0)$, \textit{i.e.} those that survive in the limit $\hbar\rightarrow 0$. Therefore, terms $O(q^n)$ with $n>0$ are quantum, while terms $O(({ J} q)^n$) are classical. Crucially, the presence of the spin compensates the transferred momentum and gives rise to classical terms with higher powers in $q$. 

At linear order in the gravitational coupling, one obtains the most general stress-energy tensor as a function of the spin and transferred momentum subject to the constrains outlined above. In \cite{Bianchi:2024shc} it was shown that such tensor, in arbitrary dimension, has the following expression:
\begin{widetext}
\begin{equation}
    T^{\mu\nu} (q) =u^\mu u^\nu m\sum_{\ell=0}^{+\infty}F_{2\ell,1}\zeta^{2\ell}    +m\sum_{\ell=1}^{+\infty}F_{2\ell,2}(S\cdot q)^\mu(S\cdot q)^\nu\zeta^{2(\ell-1)}-\frac{i}{2}mq_\rho(u^\mu S^{\nu\rho}+u^\nu S^{\mu\rho})\sum_{\ell=0}^{+\infty}F_{2\ell+1,3}\zeta^{2\ell}
    \ ,
\label{EMT}
\end{equation}
\end{widetext}
where
\begin{equation}
   S = J/m
\end{equation}
is the spin per unit mass, and we have defined 
\begin{equation}
    \zeta\equiv\sqrt{q\cdot S\cdot S\cdot q} = \sqrt{q^\mu S_{\mu}{}^{\nu}S_{\nu}{}^{\sigma}q_{\sigma}} \ .
\end{equation}
The $F$'s are constant coefficients dubbed `form factors', and the universality of the mass and dipole terms implies that $F_{0,1} = F_{1,3}=1$.

\section{Totally symmetric fields}
\label{sez 2}
We start our considerations from a massive tensorial field $\varphi^{\mu_1\dots\mu_s}(p)$ with $s$ completely symmetric indices with mass $m$ \cite{10.1098/rspa.1939.0140,PhysRevD.18.3624}.
It corresponds to a single row Young Tableau $Y(s)$
\begin{equation}
    \underbrace{\begin{ytableau}
        {} & {} & {}
    \end{ytableau}
    \dots
    \begin{ytableau}
        {} & {}
    \end{ytableau}}_s \ .
\end{equation}
For an object with $s$ symmetric indices, the spatial component of the Lorentz generator is 
\begin{widetext}
\begin{equation}
    (M^{ij})^{a_1\dots a_s, \ b_1\dots b_s}=-i\frac{(-1)^s}{s(s-1)}\sum_{r=1}^s\sum_{n\neq r}(\delta^{i(a_r}\delta^{jb_r}-\delta^{ib_r}\delta^{j(a_r})\delta^{a_n)b_n}\prod_{k\neq r,n}\delta^{a_kb_k} \ .
\end{equation}
\end{widetext}
We can identify three kinds of classical structures involving the spin tensor $S^{ij}$:
\begin{equation}
\begin{split}
    &(q_a S^{ac} S_c{}^b q_b)^\ell \ ,\\& (S^{ik}q_k)(q_aS^{ac}S_c{}^b q_b)^\ell \ , \\& (S^{ik}S^{jl}q_kq_l)(q_aS^{ac}S_c{}^b q_b)^\ell \ .
    \label{multipoli}
\end{split}
\end{equation}
For the simple case of $(q_aS^{ac}S_c{}^bq_b)$, we can see that
\begin{equation}
    (q_aS^{ac}S_c{}^bq_b)=-\frac{s^2}{m^2}\bigg(\frac{d-4}{s}+2\bigg)(q_a\epsilon_2^{ac_2\dots c_s}\epsilon_1{}_{c_2\dots c_s}{}^bq_b) \ ,
\end{equation}
which in the limit $s\gg 1$ corresponds to 
\begin{equation}
    (q_aS^{ac}S_c{}^bq_b)\approx-\frac{2s^2}{m^2}(q_a\epsilon_2^{ac_2\dots c_s}\epsilon_1{}_{c_2\dots c_s}{}^bq_b) \ .
\end{equation}
Therefore, in the high-spin limit the dimensional dependence is dropped. This is a general feature of each structure at any order. We obtain a generic dimensional independent correspondence for the three possible structures:
\begin{widetext}
    \begin{equation}
    \begin{aligned}
    &(q\cdot S\cdot S\cdot q)^\ell\xrightarrow{s\gg 1}\frac{(-1)^s}{m^{2\ell}}s^{2\ell}2^{\ell}(q^\ell\cdot \varepsilon_2\cdot \varepsilon_1\cdot q^\ell) \ , \\ \\&
    (S^{\mu\rho}q_\rho)(q\cdot S\cdot S\cdot q)^\ell\xrightarrow{s\gg 1}\frac{i}{m^{2\ell+1}}s^{2\ell+1}2^{\ell}\bigg((\varepsilon_2\cdot q^\ell)^\mu(\varepsilon_1\cdot q^{\ell+1})-(\varepsilon_2\cdot q^{\ell+1})(\varepsilon_1\cdot q^\ell)^\mu\bigg) \ ,\\  \\&
    \begin{aligned}
    (S^{\mu\rho}S^{\nu\sigma}q_\rho q_\sigma)(q\cdot S\cdot S\cdot q)^\ell\xrightarrow{s\gg 1}-\frac{(-1)^{s}}{m^{2\ell+2}}2^{\ell}s^{2\ell+2}&\bigg[\bigg((q^{\ell+2}\cdot \varepsilon_2)\cdot(q^\ell\cdot\varepsilon_1)^{\mu\nu}+(\varepsilon_2\cdot q^\ell)^{\mu\nu}\cdot(\varepsilon_1\cdot q^{\ell+2})\bigg)-\\&-\bigg((\varepsilon_2\cdot q^{\ell+1})^\mu\cdot(\varepsilon_1\cdot q^{\ell+1})^\nu+(\varepsilon_2\cdot q^{\ell+1})^\nu\cdot(\varepsilon_1\cdot q^{\ell+1})^\mu\bigg)\bigg] \ ,
    \end{aligned}
    \end{aligned}
    \label{multip}
\end{equation}
\end{widetext}
where we use the short notations in the Appendix-\eqref{Notations}.
The independence of the classical spin-tensor structures from the spacetime dimension $D$ in the high-spin limit represents an important feature. As we will see in the next subsection, this will lead to a universal form of the minimal coupling in the high-spin limit, which enforces the special symmetric structure of four dimensional spacetime. This, however, represents a general feature for any phenomenologically relevant multipolar structure. In the study of the two-body problem in gravity, higher orders in the post-Minkowskian expansion are obtained through loop computations. In order to solve the loop integrals, dimensional regularization must be implemented \cite{Akpinar:2025byi,Bohnenblust:2024hkw}. This means that we perform the continuation
\begin{equation}
    D=4\rightarrow D=4-2\epsilon \ .
\end{equation}
If we perform the computation considering finite-spin representations, such as a vector field, we obtain spin-tensor structures that acquire an explicit dependence on the spacetime dimension. In general, this could lead to finite terms, i.e. $O(\epsilon^0)$, but phenomenologically we are interested in objects in the infinite spin limit. This means that we can drop these corrections to the multipolar structure in the phenomenologically relevant regime.

\subsection{Minimal coupling}
We present the Minimal Coupling (MC) action involving massive generic $s$-symmetric indices fields, following the approach of \cite{Zinoviev:2001dt,Cangemi:2022bew}. We introduce a tower of $k=0,1,\dots,s$ symmetric tensorial fields and gauge parameters
\begin{equation}
\begin{split}
    &\varphi^{\mu[k]}:=\varphi^{\mu_1\mu_2\dots\mu_{k}} \ , \\&
    \xi^{\mu[k]}:=\xi^{\mu_1\mu_2\dots\mu_k} \ .
\end{split}
\end{equation}
The gauge parameters are traceless, $i.e.$ $Tr \ \xi^{\mu[k]}=0$. Furthermore, the fields are double traceless, i.e. $Tr(\bar{\varphi}^{\mu[k]})=0$, with $\bar{\varphi}^{\mu[k]}=Tr \ \varphi^{\mu[k]}$.
The linearized gauge transformation can be written in the following compact way
\begin{equation}
    \delta\varphi^{\mu[k]}=\partial^{(\mu_1}\xi^{\mu[k-1])}+m\alpha_{k}\xi^{\mu[k]}+m\beta_k\eta^{(\mu_1\mu_2}\xi^{\mu[k-2])} \ ,
\end{equation}
with $k$ a set of Lorentz indices. The parameters $\alpha$ and $\beta$ are
\begin{equation}
\begin{split}
    &\alpha_k=\sqrt{\frac{(s-k)(s+k+d-2)}{(k+1)(2k+d-1)}} \ , \\&
    \beta_k=\frac{k}{2k+d-5}\alpha_{k-1} \ ,
\end{split}
\end{equation}
such that the gauge parameter $\xi^{\mu[s]}$ with the highest spin is decoupled and then, $\varphi^{\mu[s]}$ has the physical degree of freedom.
The free theory $\mathcal{L}_0$ invariant under gauge transformation, can be written as a Feynman-gauge $\mathcal{L}_F$ and a sum over the gauge-fixing functions $G^{\mu[k]}$,
\begin{equation}
    \mathcal{L}_0=\mathcal{L}_F+\frac{1}{2}\sum_{k=0}^{s-1}(-1)^k(k+1)G^{\mu[k]}G_{\mu[k]} \ ,
\label{lagrangian simm}
\end{equation}
with
\begin{small}
\begin{equation}
\begin{split}
    \mathcal{L}_{F}=-\sum_{k=0}^s\frac{(-1)^k}{2}&\bigg[\varphi^{\mu[k]}(\square+m^2)\varphi_{\mu[k]}\\&-\frac{k(k-1)}{4}\bar{\varphi}^{\mu[k]}(\square+m^2)\bar{\varphi}_{\mu[k]} \bigg] \ ,
\end{split}
\end{equation}
\end{small}
and $G^{\mu[k]}$ gives the contributions off-diagonal
\begin{small}
\begin{equation}
\begin{split}
    G^{\mu[k]}&=\partial\cdot\varphi^{\mu[k+1]}-\frac{k}{2}\partial^{(\mu_1}\bar{\varphi}^{\mu[k+1])}\\&+m(\alpha_k\varphi^{\mu[k]}-\gamma_k\bar{\varphi}^{\mu[k+2]}-\delta_k\eta^{(\mu_1\mu_2}\bar{\varphi}^{\mu[k])}) \ ,
    \end{split}
\end{equation}
\end{small}
with
\begin{equation}
    \gamma_k=\frac{1}{2}(k+2)\alpha_{k+1} \ , \quad \delta_k=\frac{k(k-1)}{2(2k+d-3)}\alpha_k \ .
\end{equation}
From \cite{Cangemi:2022bew}, this Lagrangian reproduces the Kerr solution in four dimensions.
An important feature of the minimal coupling is that, at the tree level, the amplitude written with respect to the physical polarization tensors is independent of dimensions\footnote{This is true only at tree level, since at loop level we must account for both the auxiliary fields, whose couplings depend on the spacetime dimension, and the loop computations themselves.}. So when we move to arbitrary $D=d+1$, the only dependence of the dimension could come from the identification between Spin Tensor and the polarization tensors. But as we saw from \eqref{multip}, this identification is independent from the dimension when $s\gg 1$. That means that the minimal form factors generated by an higher spin object remain untouched in arbitrary dimensions, and they are precisely those which reproduce the Kerr multipolar structure in four dimensions.
From the computations at finite spin, we find that the Lagrangian \eqref{lagrangian simm} does not produce $T^{ij}$ components and, consequently, the coefficients $F_{2\ell,2}^{(MC)}$. Thus, for the fields $\varphi^{\mu_1\dots\mu_s}$, we have the following form factors \cite{Vines:2017hyw}:
\begin{small}
\begin{equation}
\begin{split}
    &F_{2\ell,1}^{(MC)}=\frac{(-1)^\ell}{(2\ell)!} \ , \quad
    F_{2\ell,2}^{(MC)}=0 \ , \quad
    F_{2\ell+1,3}^{(MC)}=\frac{(-1)^\ell}{(2\ell+1)!} \ .
\end{split}
\label{ans}
\end{equation}
\end{small}
This is perfectly in line with the fact that, in four dimensions, the only independent representations are the totally symmetric ones, and the only independent even spin orders are those encoded by the form factors $F_{2\ell,1}$, as a consequence of the relation $S^{ik}=\epsilon^{ijk}s^k$.

\subsection{Non-Minimal coupling}
We now turn to the non-Minimal Coupling (nMC). We want to introduce non-minimal operators into the Lagrangian such that the resulting stress-energy tensor reproduces the multipolar structure of the Myers-Perry (MP) solution.
The form factors of the stress-energy tensor become
\begin{equation}
\begin{split}
    &F_{2\ell,1}=F_{2\ell,1}^{(MC)}+F_{2\ell,1}^{(nMC)}=F_{2\ell,1}^{(MP)} \ ,\\&
    F_{2\ell+1,3}=F_{2\ell+1,3}^{(MC)}+F_{2\ell+1,3}^{(nMC)}=F_{2\ell+1,3}^{(MP)} \ ,\\&
    F_{2\ell,2}=F_{2\ell,2}^{(MC)}+F_{2\ell,2}^{(nMC)}=F_{2\ell,2}^{(MP)}\ .
\end{split}
\end{equation}
The total action consists of a Minimal and a non-Minimal contribution,
\begin{equation}
    S=\int d^Dx\sqrt{-g}\bigg(\mathcal{L}_0+\mathcal{L}_{nMC}\bigg) \ .
\end{equation}
The most general non-minimal Lagrangian contributing to the deformation of the classical spin orders, up to couplings to the auxiliary fields needed to preserve the gauge invariance, is given by
\begin{widetext}
\begin{equation}
\begin{split}
    \mathcal{L}_{nMC}&=\sum_{\ell=1}^{s}(-1)^ss^{2\ell}\frac{C_{nMC}^{(1),\ell}}{m^{2\ell}}\bigg(D^{\beta_1}\dots D^{\beta_{\ell-1}}D^{\beta'_1}\dots D^{\beta'_{\ell-1}}R_{\mu\nu\rho\sigma}\bigg)(D^\mu\varphi^\nu{}_{\beta_1\dots\beta_{\ell-1}\alpha_{\ell}\dots\alpha_{s-1}})(D^{\rho}\varphi^\sigma{}_{\beta'_1\dots\beta'_{\ell-1}}{}^{\alpha_{\ell}\dots\alpha_{s-1}})\\ \\ &
    +\sum_{\ell=1}^{s-1}s^{2\ell+1}\frac{C_{nMC}^{(3),\ell}}{m^{2\ell}}\bigg(D^{\beta_1}\dots D^{\beta_{\ell-1}}D^{\beta'_1}\dots D^{\beta'_{\ell-1}}D_\lambda R_{\mu\nu\rho\sigma}\bigg)(D^\mu\varphi^{\nu\rho}{}_{\beta_1\dots\beta_{\ell-1}\alpha_{\ell}\dots\alpha_{s-2}})(\varphi^{\sigma\lambda}{}_{\beta'_1\dots\beta_{\ell-1}'}{}^{\alpha_{\ell}\dots\alpha_{s-2}})\\ \\&
    +\sum_{\ell=1}^{s-1}(-1)^ss^{2\ell}\frac{C_{nMC}^{(2),\ell}}{m^{2\ell-2}}\bigg(D^{\beta_1}\dots D^{\beta_{\ell-1}}D^{\beta'_1}\dots D^{\beta'_{\ell-1}}R_{\mu\nu\rho\sigma}\bigg)(\varphi^{\mu\rho}{}_{\beta_1\dots\beta_{\ell-1}\alpha_{\ell}\dots\alpha_{s-2}})(\varphi^{\nu\sigma}{}_{\beta'_1\dots\beta_{\ell-1}'}{}^{\alpha_{\ell}\dots\alpha_{s-2}}) \ .
\end{split}
\label{non minimal}
\end{equation}
\end{widetext}
The non-minimal Lagrangian in \eqref{non minimal} gives rise to the following contribution to the stress-energy tensor:
\begin{widetext}
\begin{equation}
\begin{split}
    T^{\mu\nu}_{(nMC)}&=\sum_{\ell=1}^{s}\frac{m}{4}\frac{C^{(1),\ell}_{nMC}}{2^{\ell}}u^\mu u^\nu(q\cdot S\cdot S \cdot q)^{\ell}
    -\sum_{\ell=1}^{s-1}\frac{im}{4}\frac{C^{(3),\ell}_{nMC}}{2^{\ell}}\bigg(u^\mu(S\cdot q)^\nu+u^\nu(S\cdot q)^\mu\bigg)(q\cdot S\cdot S\cdot q)^{\ell}-\\\\&
    -\sum_{\ell=1}^{s-1}\frac{m}{4}\frac{C_{nMC}^{(2),\ell}}{2^{\ell-1}}(S\cdot q)^\mu (S\cdot q)^\nu (q\cdot S\cdot S\cdot q)^{\ell-1} \ ,
\end{split}
\end{equation}
\end{widetext}

Therefore, the total stress-energy tensor is characterized by the following form factors, which we require to match those of the Myers-Perry solution
\begin{equation}
    \begin{cases}
        F_{2\ell,1}^{(MC)}+\frac{C_{nMC}^{(1),\ell}}{2^{\ell+2}}=F_{2\ell,1}^{(MP)}\\\\
        F_{2\ell+1,3}^{(MC)}+\frac{C_{nMC}^{(3),\ell}}{2^{\ell+1}}=F_{2\ell+1,3}^{(MP)}\\\\
        F_{2\ell,2}^{(MC)}-\frac{C_{nMC}^{(2),\ell}}{2^{\ell+1}}=F_{2\ell,2}^{(MP)} \ .
    \end{cases}
\end{equation}
We now use the relation between the form factors in Ref. \cite{Bianchi:2024shc}
\begin{equation}
    F_{2\ell,1}^{(MP)}=F_{2\ell,2}^{(MP)}+F_{2\ell+1,3}^{(MP)} \ .
\end{equation}
Then, using the Legendre duplication formula, we obtain the following relation among the non-minimal coefficients:
\begin{equation}
C^{(1),\ell}_{nMC}+2C^{(2),\ell}_{nMC}-2C^{(3),\ell}_{nMC}=(-1)^{\ell+1}\frac{4\sqrt{\pi}}{2^\ell \Gamma(\ell)\Gamma(\ell+\frac{3}{2})} \ .
\label{rel forma 2}
\end{equation}
These relations provide dimension-independent constraints on the non-minimal coupling coefficients required to reproduce the Myers-Perry multipolar structure, in arbitrary dimensions, for an object with infinite spin.
\section{Mixed-Symmetric Fields}
\label{sez 3}
In this section, we extend our analysis to generic representations. We present some starting points for future generalizations of the stress-energy tensor generated by a general Young Tableau field coupled to gravity.
In particular, here we focus on the simplest extension of one family of symmetric indices, which is the two-rows Young Tableau $Y(s,n)$:
\begin{equation}
\begin{split}
    \ \overbrace{\underbrace{\begin{ytableau}
        {} & {}  \\
        {} & {}  
    \end{ytableau}\dots\begin{ytableau}
        {}  \\
        {} 
    \end{ytableau}}_{n}\dots \begin{ytableau}
        {} & {} 
    \end{ytableau}}^{s} \ .
\end{split}
\label{representation}
\end{equation}
It corresponds to a field with two families of symmetric indices, $\varphi_{\mu_1\dots\mu_s,\nu_1\dots\nu_n}$, subject to the Young constraint
\begin{equation}
    \varphi_{(\mu_1\dots\mu_s,\nu_1)\dots\nu_n}=0 \ .
\end{equation}
For this field, the spin operator is given by
\begin{equation}
    (M^{ij})^{a^1[s] a^2[n],b^1[s]b^2[n]} \ ,
\end{equation}
where $a^i,b^i$ denote sets of indices belonging to the $i$-th family.
If we consider at least $s\gg 1$, the identification of the spin-tensor structures becomes independent of the dimension of spacetime. This result can be generalized to an arbitrary Young Tableau representation $Y(s_1,\dots,s_k)$ with $s_1>s_2>\dots>s_k$, corresponding to fields with $k$ families of symmetric indices subject to the Young constraint. In this case, the structures in \eqref{multipoli} become independent of the spacetime dimension when at least $s_1\gg 1$.

\subsection{Minimal Coupling}
The general free Lagrangian for a field described by a two rows Young Tableau $Y(s,n)$ \cite{Reshetnyak:2026lno} can be written as 
\begin{equation}
    \mathcal{L}=\frac{(-1)^{s+n}}{2}\bigg(\varphi_{\mu[s],\nu[n]}(x)(\square+m^2)\varphi^{\mu[s],\nu[n]}(x)+\mathcal{L}_{GF}\bigg) \ ,
\end{equation}
with $\mu[s]=\mu_1\dots\mu_s$ and $\nu[n]=\nu_1\dots\nu_n$ are sets of indices belonging to the same family. Here, the $\mathcal{L}_{GF}$ accounts for the couplings with the auxiliary fields and the gauge-fixing terms. In this case, the minimal stress-energy tensor is more difficult to generalize, since the Young constraint depends on the number of indices in each of the two families. For this reason, we leave the generalization of the minimal coupling to fields with arbitrary Young Tableau representations for future work. Nevertheless, some insights can be obtained from the results of \cite{Campanella:2026wqt}, where an antisymmetric tensor coupled to gravity in four dimensions was shown to generate the same form factors that reproduce the multipolar structure of Kerr. We may therefore identify a subset of fields that reduce to unconstrained totally symmetric fields, as studied in Section \eqref{sez 2}, and exhibit the same multipolar structure in four dimensions and, consequently, in arbitrary dimensions in the high-spin limit.\footnote{ For instance, consider a field $\varphi_{[\mu\nu],\rho}$ (See Ref. \cite{zinoviev2002massivemixedsymmetrytensor} for its specific Lagrangian). It is dual to a spin-2 field in four dimensions.} At present, however, this remains a conjecture based on previous results.

\subsection{Non-Minimal coupling}
Finally, we show how the non-minimal operators required to modify the multipolar expansion become more involved for mixed-symmetry fields. We focus in particular on the quadratic order in spin, considering the regime in which both $s$ and $n$ are much larger than spacetime dimension. In this limit, we consider the following non-minimal operators:
\begin{widetext}
\begin{equation}
\begin{split}
    \Delta\mathcal{L}_1=8C_1\frac{(-1)^{s+n}}{m^2}R_{\mu\nu\rho\sigma}&\bigg(s^2(D^{\mu}\varphi^{\nu \alpha[s-1],\beta[n]})(D^\rho\varphi^\sigma{}_{\alpha[s-1],\beta[n]})+n^2(D^\mu\varphi^{\alpha[s],\nu\beta[n-1]})(D^\rho\varphi_{\alpha[s],}{}^\sigma{}_{\beta[n-1]})\\&
    +2sn(D^\mu\varphi^{\nu\alpha[s-1],\lambda\beta[n-1]})(D^\rho\varphi_{\lambda\alpha[s-1],}{}^\sigma{}_{\beta[n-1]})\bigg) \ ,
\end{split}
\label{op1}
\end{equation}
\end{widetext}
and
\begin{widetext}
\begin{equation}
\begin{split}
    \Delta\mathcal{L}_2=4C_2(-1)^{s+n}R_{\mu\nu\rho\sigma}&\bigg(s^2\varphi^{\mu\rho \ \alpha[s-2],\beta[n]}\varphi^{\nu\sigma}{}_{\alpha[s-2],\beta[n]}+n^2\varphi^{\alpha[s],\mu\rho \ \beta[n-2]}\varphi_{\alpha[s],}{}^{\nu\sigma}{}_{\beta[n-2]}\\&
    +2sn\varphi^{\mu\alpha[s-1],\rho\beta[n-1]}\varphi^\nu{}_{\alpha[s-1],}{}^{\sigma}{}_{\beta[n-1]}\bigg) \ ,
    \end{split}
    \label{op2}
\end{equation}
\end{widetext}
where $C_1$ and $C_2$ are dimensionless coupling constants. These expressions show the increased complexity of the non-minimal couplings required to generate higher-order multipole structures for mixed-symmetry fields. In particular, one must account for the different possible contractions involving the distinct families of indices. More generally, for a representation $Y(s_1,\dots,s_k)$, one expects $k(k+1)/2$ classes of contractions between the index families. However, the operators involving mixed contractions are not necessarily independent, owing to the Young constraint.
The operators in \eqref{op1}-\eqref{op2} deform the stress-energy tensor as follows:
\begin{small}
\begin{equation}
    \Delta T^{\mu\nu}=m\bigg(-C_1u^\mu u^\nu (q\cdot S\cdot S\cdot q)+C_2(S^{\mu\rho}S^{\nu\sigma}q_\rho q_\sigma)\bigg) \ .
\end{equation}
\end{small}

\section{Discussion and Outlook}
\label{sez 4}
We have studied the stress-energy tensor generated by higher-spin fields in various Lorentz representations coupled to gravity in arbitrary spacetime dimension. Motivated by the results of \cite{Campanella:2026wqt}, where spin universality is lost in dimensions greater than four, we investigated whether a limit exists in which a form of universality is restored in arbitrary dimensions.
We first considered fields with a single family of $s$ symmetric indices. By taking the high-spin limit, we showed that the spin-tensor structures can be identified through dimension-independent combinations of the polarization tensors. As a result, the form factors associated with the minimal coupling remain unchanged in arbitrary dimensions and they are those which reproduce the Kerr multipolar structure in four dimensions. This feature is particularly relevant for dimensional regularization in loop computations, as dimensional corrections to the spin-tensor structures can be neglected in the physically relevant high-spin regime. Furthermore, exploiting this dimensional independence, we derived dimension-independent constraints on the non-minimal couplings required to reproduce the Myers-Perry multipolar structure. We then extended our analysis to mixed-symmetry fields, focusing on two rows representations corresponding to two families of symmetric indices subject to the Young constraint. We showed that the spin-tensor structures become independent of the spacetime dimension when at least one of the spin quantum number is taken to the high-spin limit. This result can be generalized to arbitrary Young Tableau representations with $k$ families of indices, provided that at least one of the spin quantum numbers is taken to be large. Finally, we provided some insights into the non-minimal coupling and showed how the form of the operators required to deform the multipolar structures becomes increasingly involved as the number of index families grows. These considerations provide starting points for future generalizations of the stress-energy tensor generated by fields in arbitrary Young tableau representations. 
Another possible direction is to consider charged higher-spin fields, extending the results of \cite{Gambino:2025iyx} for the gravitational metric and the gauge potential. Future work could also investigate the structure of non-minimal couplings for rotating objects other than Myers-Perry, in the high-spin limit. More generally, alternative approaches to probing the multipolar structure of rotating objects could be explored, such as the scattering of massive probes off gravitational backgrounds in the Kerr-Schild gauge, following the recent work on Myers-Perry black holes in \cite{Bianchi:2025xol}.
Finally, it would be interesting to explore whether the results obtained in this work can be reproduced from scattering amplitudes involving highly excited string states, in the spirit of recent works on the emergence of Kerr-like multipoles from string theory \cite{Alessio:2025nzd,Firrotta:2026cvk}. 

\begin{acknowledgments}
I would like to thank Fabio Riccioni for the useful discussion and suggestion on the paper. I would like to thank also Andrea Campoleoni, Dario Francia and Yurii Zinoviev for the discussions and references about Higher Spin and Mixed Symmetric fields, as well as Massimo Bianchi for reading and suggestions over the manuscript. Finally, I would like to thank Carlo Heissenberg and the Institut de Physique Théorique (Ipht) of Saclay for the hospitality in the last stages of this work. This work is partially supported by the INFN TEONGRAV initiatives.
\end{acknowledgments}

\appendix
\begin{widetext}
\section{Notations}
\label{Notations}
We work in mostly negative signature with $\eta_{00}=+1$ and in natural units, $\hbar=c=1$, whereas we keep the gravitational coupling constant $G$ explicit. Greek indices $\mu,\nu=0,1,...,d$   are for spacetime components and the Latin indices are for space components only $i,j=1,...,d$, where $D=d+1$ is the number of spacetime dimensions. Repeated Latin indices are always meant to be contracted by $\delta^{ij}$, and there is no sign difference between upstairs and downstairs indices.
\newline
For the multipole structures in \eqref{multip}, we use the short notations for $2\ell$ orders
\begin{equation}
    q^\ell\cdot\varepsilon_2\cdot\varepsilon_1\cdot q^\ell=q_{\mu_1}\dots q_{\mu_\ell}\varepsilon_2^{\mu_1\dots \mu_\ell \lambda_{\ell+1}\dots \lambda_{s}}\varepsilon_1{}_{\lambda_{\ell+1}\dots \lambda_s}{}^{\nu_1\dots \nu_\ell}q_{\nu_1}\dots q_{\nu_\ell} \ ,
\end{equation}
for the $2\ell+1$ orders
\begin{equation}
    (\varepsilon_2\cdot q)\cdot\varepsilon^\mu_1=q_\rho\varepsilon_2^{\rho \lambda_1\dots \lambda_{s-1}}\varepsilon_1{}^\mu{}_{\lambda_1\dots \lambda_{s-1}} \ ,
\end{equation}
and for the $2\ell+2$ orders
\begin{equation}
    \begin{split}
        &q^2\cdot\varepsilon_2\cdot\varepsilon_1^{\mu\nu}=q_{\rho_1}q_{\rho_2}\varepsilon_2^{\rho_1\rho_2\lambda_1\dots \lambda_{s-2}}\varepsilon_1^{\mu\nu}{}_{\lambda_1\dots \lambda_{s-2}}\\&
    \varepsilon_2^{\mu\nu}\cdot\varepsilon_1\cdot q^2=\varepsilon_2^{\mu\nu \lambda_1\dots \lambda_{s-2}}\varepsilon_1^{\rho_1\rho_2}{}_{\lambda_1\dots \lambda_{s-2}}q_{\rho_1}q_{\rho_2}\\&
    (\varepsilon_2\cdot q)^\mu(\varepsilon_1\cdot q)^\nu=q_{\rho_1}q_{\rho_2}\varepsilon_2^{\mu \rho_1\lambda_1\dots \lambda_{s-2}}\varepsilon_1^{\nu \rho_2}{}_{\lambda_1\dots \lambda_{s-2}} \ .
    \end{split}
\end{equation}

\end{widetext}

\bibliography{biblio.bib}

\end{document}